 \documentclass[aps,prl,twocolumn,groupedaddress]{revtex4}

\usepackage{graphicx}
\usepackage{dcolumn}
\usepackage{bm}

\begin{document}


\title{
 Core-excitation three-cluster model description of  $^8$He and $^{10}$He 
}

\author{
H. Kamada
}
\email{kamada@mns.kyutech.ac.jp}
\affiliation{
Department  of Physics, Faculty of Engineering, Kyushu Institute of Technology,
 1-1 Sensuicho, Tobata, Kitakyushu 804-8550, Japan
}

\author{
 M. Yamaguchi
}
\email{yamagu@rcnp.osaka-u.ac.jp}
\affiliation{
Research Center for Nuclear Physics, Osaka University,
Ibaraki 567-0047, Japan
}

\author{
E. Uzu
}
\email{uzu@ee.kagu.tus.ac.jp}
\affiliation{
Liberal Arts, Faculty of Engineering Division 2,
Tokyo University of Science,
1-3 Kagurazaka, Shinjuku-ku Tokyo 162-8601, Japan
}

\date{\today}

\begin{abstract}
We introduce a new model applying to the core-nucleus and two-neutron system. 
The Faddeev equations of
$^6$He-n-n and $^8$He-n-n systems for $^8$He and $^{10}$He are solved, respectively. 
The potential of the subsystem in the
model has been determined to make a coupling both of the ground state and the excited one inside the core
nucleus. By a similar mechanism the three-nucleon system is solved with the three-body force originating from
an isobar excitation of the nucleon. Inputting only the information of subsystem energy levels and widths we
get the coupling constants of rank 1 Yamaguchi potential between the core nucleus and neutron. We calculate
the Faddeev three-cluster equations to obtain the low-lying energy levels of 
$^8$He and $^{10}$He. The 1$^-$Â state of $^{10}$He,
which has not been detected yet in experiments, 
is located in the energy level between the 0$^+$ and 2$^+$ states.
\end{abstract}

\pacs{27.20.+n, 21.45.-v, 21.60.Gx}

\maketitle

\section{I. INTRODUCTION}

Due to developments of experimental technique, our
knowledge of unstable nuclei has been increasing rapidly.
Experimental researchers have recently reported a lot of
events. Here neutron-rich nuclei are good targets for studying
interesting phenomena, e.g., clustering, halos, deformation,
dineutron correlation, etc. In order to look for these properties
which differ from ordinary shell model study, one may need to
employ cluster model calculations. However, the interactions
between clusters are usually very complex, except for the
$\alpha$ cluster model treated as the resonating group method.
According to ab initio calculations, there are at most four-body
calculations\cite{Kamada:2001tv}. Four-nucleon scattering has been solved
by the Faddeev-Yakubovsky formalism using the realistic
nucleon-nucleon force including the three-body force\cite{Viviani:2011ax}.
Beyond the four-nucleon system there are computational
difficulties because of limited memory size and CPU time.
Nevertheless, the Green's function Monte Carlo simulation is
very promising. Recent calculations show many energy spectra
up to A = 9 \cite{Pastore:2013ria}.

There are some microscopic or effective theoretical approaches.
For instance, the cluster orbital shellmodel (COSM),
complex scaling method (CSM)\cite{Aguilar:1971ve}, and the method of analytic
continuation in the coupling constant (ACCC)\cite{Kukulin:1979qn} describe $^9$He
and $^{10}$He nuclei by their core-nucleus + valence-neutrons
model \cite{Aoyama:1997zz,Aoyama:2002zz}. Systematic studies from $^5$He to $^8$He are reported
on the basis of the tensor-optimized shell model (TOSM)
\cite{Myo:2011ci} using a bare nucleon-nucleon interaction, of which the
short-range correlation is treated by the unitary correlation
operator method (UCOM) \cite{Neff:2002nu}.

On the other hand, the three-cluster model of the  Faddeev theory 
has been applied to the low-lying energy states of the $^6$Li nucleus 
as $\alpha$ + n + p three-body system using nonlocal separable interactions
\cite{Eskan1992}.
In the case of T=1 the isotope $^6$He 
the binding energy and widths of the resonance for 
the ground state J$^\pi$=0$^+$ 
and the  resonance state J$^\pi$=2$^+$ agree with experiment.
By the same scheme we have also  been  investigating  other exotic nucleus $^{9}_\Lambda$Be
of $\alpha$ + $\alpha$ + $\Lambda$ three-body system\cite{Oryu2000,Cravo:2002jv}.

In the next section we will introduce a new model calculation based on the 
Faddeev theory. The three-body system is treated as the cluster model 
consisting of core-nucleus + n + n to investigate $^8$He and $^{10}$He 
nuclei. 
$^{6,8}$He are so-called Borromean nuclei and 
 $^{10}$He is also regarded as the Borromean nucleus
 because the energy level 
of the ground state is much closer  to
the three-body breakup threshold.  
It is often considered that 
the core-nucleus of the three-body model deals with only  
the ground state core-nucleus. However, 
in our model not only the ground state core-particle but also
an  excited state core-nucleus are adopted.
The idea \cite{Koike:2003}  is also found in the case of the  3-nucleon system, 
in which some of nucleons become delta isobar in $^3$He\cite{Pena:1990ea}.    

Preliminary calculations have been carried out
\cite{Uzu2007,Yamaguchi:2007}. 
Because the excited state J$^\pi$=${1\over2}^-$ of $^7$He  was  not found 
in the experiment, in the former work 
$^8$He ground state coluld not be described accurately.
 Using the presence of the excited state in the experiment\cite{Skaza:2006cc}
 we recalculate with the new data of $^7$He.   Our theoretical 
prediction will be demonstrated 
 in case of $^8$He and $^{10}$He nuclei in section 3.
The conclusion is  given in section 4.

\section{II. A new three-cluster model}

In the framework of the Faddeev theory the three-body equations were 
represented 
as the Alt-Grassberger-Sandhas (AGS) equations using a separable potential 
of NN interaction\cite{Lovelace:1964mq}.
The AGS equations are used in many three-body systems. 
It has succeeded in calculation of a
  three-body breakup process for  the $\alpha$- n -p system 
first \cite{Koike:1978}. Recently the study of the 
system has progressed well 
\cite{Deltuva:2006um}. 
The system was often investigated and the calculation of the resonance 
states T=1 without Coulomb force are
discussed just corresponding to the case of $^{6}$He nucleus. 
We verified the former work \cite{Eskan1992} and 
the energy level of the ground state 
J$^\pi$=0$^+$ from the threshold of $ \alpha$ +2n is obtained as 
-- 0.56 MeV  vs. data -- 0.973 MeV.
The energy level of the first excited state J$^\pi$=2$^+$ 
is also obtained as 0.95 MeV ( $\Gamma$ =0.3MeV) vs. 
data 0.824MeV ( $\Gamma$ =0.113MeV)\cite{Tilley:2002vg}.    
The separable potential is very primitive, nevertheless, these calculations 
encourage us to start  
neutron-rich study. 

On the other hand, the research in three-nucleon scattering has 
made great progress according to 
the three-body force\cite{Gloeckle:1995jg,Witala:2011yq}. 
It is considered that the fundamental origin of the three-body forces comes
from the delta excitation, or inner excitation, of 
nucleon\cite{Fujita:1957zz}. 
Study of the three-nucleon force 
is progressing recently centering 
on the chiral symmetry which QCD Lagrangian 
possesses\cite{Epelbaum:2007sq,Epelbaum:2013hya}. 

\begin{figure}[htbp]
\begin{center}
\includegraphics[scale=0.25,angle=0]{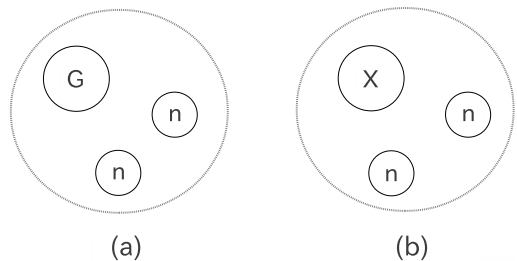}
\end{center}
\caption{Illustration of core-excitation cluster. The core-cluster
 of the ground state and excited state   
are labeled "G" and "X", respectively. Neutrons are also labeled "n".    }
\label{Fig1}
\end{figure}

If the idea of the inner excitation is applied to 
the case of neutron-rich nuclei, 
more precise theoretical expectations would be possible taking into 
 consideration  the inner excitation of the core-cluster which 
constitutes the nucleus\cite{Koike:2003}.  This idea has a similarity 
to the  delta  isobar excitation 
in the three-nucleon system \cite{Pena:1990ea}. 
Illustrations of the model which we imagine, are shown in Fig. \ref{Fig1}. 
Labels "G" of Fig. 1 (a) and "X" of Fig. 1 (b) mean the 
names of the ground state core-nucleus and the 
excited state one, respectively.

The Hilbert space $\cal H$ of the model consists of two Hilbert ones 
;

\begin{eqnarray}
{\cal H}={\cal H}(G) \textcircled{+}{\cal H}(X).
\end{eqnarray}
Using the word of wave function, we have
\begin{eqnarray}
|\Psi \rangle = |G \rangle | \Psi_G \rangle + |X\rangle |\Psi_X\rangle ,
\end{eqnarray}
where $|G \rangle $ and $ | X \rangle $ are orthonormal basis to distinguish 
their spaces,
\begin{eqnarray}
\langle G | G \rangle  =\langle X |X \rangle =1,~~~~ 
\langle G | X \rangle =\langle X | G \rangle =0.
\end{eqnarray}
The free Hamiltonian $\hat H_0^{2clust.}$ of the  subsystem consisting of 
the core-nucleus and neutron is represented as
\begin{eqnarray}
&&\hat H_0^{2clust.} |G \rangle \equiv {p^2 \over 2 \nu } | G \rangle ,
\cr
&& \hat H_0^{2clust.} | X \rangle \equiv (\delta m + {p^2 \over 2\nu } )| X \rangle,
\end{eqnarray}
where $p$ and $\nu$ are the relative momentum and the reduced  mass between the 
core-nucleus and neutron, respectively. The mass difference $\delta m$ 
is the energy level shift of  the ground core-nucleus and the excited one.

\subsection{Two-Body interaction}

In our model the potential of two-cluster system
has  a rank 1 separable Yamaguchi form using a simple formfactor $g(p)$. 
For instance,
the neutron-neutron potential of $^1$S$_0$ partial wave is given as 
\begin{eqnarray}
V_{nn}(p,p')=-\gamma_{nn}^2 g_{nn}(p)g_{nn}(p')
\end{eqnarray}
with
\begin{eqnarray}
g_{nn}(p)={1\over p^2+\beta_{nn}^2},
\end{eqnarray}
where we choose parameters as 
$\beta_{nn}$=1.1648 fm$^{-1}$ and $\gamma_{nn}^2=$0.3943fm$^{-3}$
 from \cite{Eskan1992}.

Let us introduce a new form factor $h$, which is combined
 with the partial waves
$| l_I S_I j_I \rangle $ and the particle  basis $| I \rangle$;
\begin{eqnarray}
\langle p | h \rangle =\sum_{I=G,X} \sum_{l_I, S_I, j_I} 
\gamma_{In;l_I,S_I,j_I}~ g_{In;l_I,S_I,j_I}(p) | l_I S_I j_I  \rangle | I \rangle
\cr &&
\label{h-func}
\end{eqnarray}
with
\begin{eqnarray}
g_{In;l_I,S_I,j_I}(p)={p^{l_I}\over(p^2+\beta_{In;l_I,S_I,j_I}^2)^{l_I+1}}
\end{eqnarray}
where $l_I$, $S_I$ and $j_I$ are angular momentum, total spin  
and total angular momentum of 2-body subsystem ($j_I = l_I+S_I$),
respectively. 
 The core-nuclei neutron potential $V$ is given by  the formfactor $h$,
\begin{eqnarray}
\hat V = - | h \rangle \langle h |.
\end{eqnarray}
However, the neutron-neutron (nn) potential $\hat V_{nn}$differs from 
this form, 
one writes
it as
\begin{eqnarray}
\hat V_{nn}=-|g_{nn} \rangle \gamma_{nn}^2
\langle g_{nn} | \{ |G\rangle \langle G| + |X\rangle \langle X |\}.
\end{eqnarray} 
Apparently  the potential $\hat V_{nn}$ is not coupled between $|G\rangle$ 
and $| X \rangle$. 

When 
the core-nucleus spin has the ground state 0$^+$ 
and the excited state 2$^+$,
there are $S_G={1\over2}$ and,  $S_X={3\over 2}$ and ${5\over 2}$, 
respectively.
If one takes the same number for the parameter $\beta$ the potentials of 
$S_X={3\over2}$ and $S_X={5\over2}$ differ only in the coupling constants.
The degenerated coupling constant $\gamma_{In;l_I,j_I}^2$ could be introduced;  
\begin{eqnarray}
&&\gamma_{Gn;l_G,j_G}^2 \equiv \gamma_{Gn;l_G,{1\over2},j_G}^2,\cr
&&\gamma_{Xn;l_X,j_X}^2 \equiv \gamma_{Xn;l_X,{3\over2},j_X}^2
+\gamma_{Xn;l_X,{5\over2},j_X}^2.
\label{sumgamma}
\end{eqnarray}
According to the separable scheme the t-matrix $t(p,p';E_2)$
\begin{eqnarray}
&&t _{In;l_I,S_I,j_I,I'n;l_I',S_I',j_I'}(p,p';E_2) 
\cr
&&
\equiv 
\langle I | \langle l_I S_I j _I |
 \langle p|h\rangle \tau (E_2) \langle h | p' 
\rangle
| l_I' S_I' j_I' \rangle
 | I' \rangle 
\end{eqnarray}
 fulfills the Lippmann-Schwinger equation
with resulting  
\begin{eqnarray}
\tau (E_2) = -1 - \tau (E_2) \langle h | \hat G_0^{2clust.}(E_2)
| h \rangle. 
\label{eq13}
\end{eqnarray}
In order to determine these coupling constants $\gamma$ in Eq. (\ref{h-func})
 we introduce 
the  following natural  assumption.
If the subsystem has no bound state (Borromean nuclei is 
just in this case) but has some resonance states, 
the propagator $\tau (E_2)$ must be diverged
 at the resonance energy $E_2=E_2^{res}$ which has a real part $E_2^{(r)}
$ and width $\Gamma$. 
Under the condition $\tau (E_2) =\infty$ Eq. (\ref{eq13}) becomes 
\begin{eqnarray}
&&1+ \gamma_{Gn;l_G,j_G}^2 \langle g _{Gn;l_G,j_G} |
{1 \over E_2^{res} -\hat p^2/2\nu +i\epsilon} |  g _{Gn;l_G,j_G} \rangle 
\cr
&&
+\gamma_{Xn;l_X,j_X}^2 \langle g _{Xn;l_X,j_X} |
{1 \over E_2^{res} -\delta m -\hat p^2/2\nu +i\epsilon} |  g _{Xn;l_X,j_X} \rangle 
\cr &&
=0
\label{cond}
\end{eqnarray}
Approximately the resonance state occurs only 
two channel and there is assumed to be no absorption channel,  
we expect these coupling constants are a real number.
Consequently, the condition leads to 2 conditions 
(real part and imaginary one) to subtract 2 unknown parameters $\gamma_G$ and $\gamma_X$.

\begin{figure}[htbp]
\begin{center}
\includegraphics[scale=0.25,angle=0]{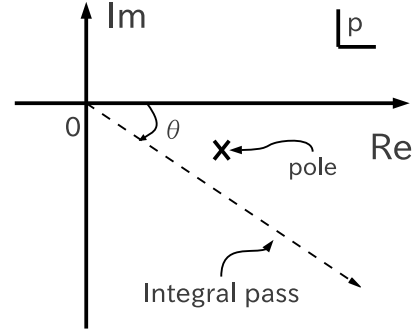}
\end{center}
\caption{Integral pass  of Eq. (\ref{cond}). In Riemann complex sheet of 
the variable $p$  
the integral pass is taken as the dashed line below the resonance 
pole.}
\label{Fig2}
\end{figure}
As shown in Fig. \ref{Fig2} one needs to take the integral pass of 
Eq. (\ref{cond}), 
 because the resonance pole is located on physical Riemann sheet 
at $p=p_{pole}$ with 
$p_{pole}^2 =2 \nu E_2^{res}$. 

In order to apply these potential to the three-body system,
 we must resolve the  degeneracy of $S_X$. 
Following a natural way of thinking the weight of the couplings will be taken from 
the degree of multiplicity under the condition of (\ref{sumgamma})
\footnote{Caution that the case $(l_X,{5\over2}, j_X) =(1,{5\over 2},{1\over2})
$ does not occur, therefore, one needs not the renormalization for
$(1,{3\over 2},{1\over2})$. },
\begin{eqnarray}
\gamma_{Xn;l_X,S_X,j_X} =\sqrt{2 S_X+1\over 10 } \gamma_{Xn;l_X,j_X}.
\end{eqnarray}

We will show these coupling constants of $^6$He-n and $^8$He-n in section III.

\subsection{Three-body integral equation}

The AGS equations are well-established \cite{Afnan:1974}, therefore, we will not repeat
the same part of Ref. \cite{Eskan1992}. The following explanation is an 
additional part 
because of the extension of core-excitation channel ($G$ or $X$) and the 
definition of the wave function.

The total wave function $| \Psi^{J^\pi T} \rangle$ with the total angular
momentum $J$, the parity $\pi$, and total isospin $T$
consists of the Faddeev components $\psi ^{J^\pi T}$ labeled by   
 particle-channel $\alpha$, $\beta$ and $\gamma$;
\begin{eqnarray}
|\Psi^{J^\pi T} \rangle = |\psi_\alpha^{J^\pi T} \rangle
                        + |\psi_\beta^{J^\pi T} \rangle
                        + |\psi_\gamma^{J^\pi T} \rangle.  \label{eq:fadcom}
\end{eqnarray}
The AGS equations for the Faddeev component  is 
given by
\begin{eqnarray}
|\psi_\alpha^{J^\pi T} \rangle &=& G_0 t_{\alpha}
                         \sum_{\beta \not= \alpha}
                         |\psi_\beta^{J^\pi T} \rangle  \label{eq:fadeq1} \\
                     &=& G_0 |h_{\alpha} \rangle
                         \tau_{\alpha} 
                         \langle h_{\alpha} |
                         \sum_{\beta \not= \alpha}
                         |\psi_\beta^{J^\pi T} \rangle . \label{eq:fadeq2}
\end{eqnarray}
The reduced wave function $f_{I;\tilde K_\alpha}^{J^\pi T} (q_\alpha)
$ is defined by
\begin{eqnarray}
&&\sum _{I=G,X}\langle I |\langle \tilde K_{\alpha} | \langle q_\alpha | f_{\alpha} ^{J^\pi T} \rangle = 
\langle q_\alpha | f_{\tilde K_\alpha}^{J^\pi T} \rangle 
= f_{\tilde K_\alpha}^{J^\pi T} (q_\alpha)
\cr &&
\equiv \sum_{I=G,X}
\gamma_{In;l_\alpha,j_\alpha} \langle g_{In;l_\alpha {j}_\alpha} |
                            \sum_{\beta \not= \alpha}
                            | \psi_{I;\beta}^{J^\pi T} \rangle, 
\end{eqnarray}
where $q_\alpha $ is the Jacobi momentum designating the momentum of the 
particle labeled by $\alpha$ relative to the ($\beta \gamma$) pair.
The index $K_\alpha = \{ t_\alpha, j_\alpha, S_\alpha, l_\alpha,
{\cal S}_\alpha, {\cal L}_\alpha, I \}$ is defined as
the quantum numbers that label the different three-body channels 
J$^\pi$ T. 
The index $\tilde K_\alpha=\{ t_\alpha, j_\alpha, l_\alpha,
{\cal S}_\alpha, {\cal L}_\alpha \}$  is also defined because of the 
degeneration of $S_\alpha$ and $I$.

Here, for the sake of unifying the notation the related  coupling constant 
$\gamma_{nn}$ is also written as $\gamma_{In;l_\alpha,j_\alpha}$ when
the spectator of the particle channel $\alpha $ is the core-nucleus.
The 
following angular momentum and isospin coupling scheme is given as
\begin{eqnarray}
&&S_\alpha = s_\beta+s_\gamma, ~~~ j_\alpha = l_\alpha + S_\alpha, ~~~ 
t_\alpha = \tau_\beta +\tau_\gamma, \cr
 && {\cal S}_\alpha = j_\alpha + s_\alpha, ~~~ J={\cal L}_\alpha + {\cal S}_\alpha, ~~~ T=t_\alpha + \tau_\alpha.
\end{eqnarray}
Here, $s_\beta$ and $\tau_\beta$ refer to the spin and isospin of the particle
labeled by $\beta$, $l_\alpha$ refers to the relative orbital angular 
momentum of the ($\beta\gamma$) pair, ${\cal S}_\alpha$  is the channel spin;
and ${\cal L}_\alpha $ is the orbital angular momentum of the spectator 
particle $\alpha$ relative to the ($\beta \gamma$) pair.

The AGS equations (\ref{eq:fadeq2}) 
are modified into equations for the  reduced wave functions;
\begin{eqnarray}
&&f_{\tilde K_\alpha}^{J^\pi T} (q_\alpha) \cr
&&= \sum_{I,I'}\sum_{\tilde K_\gamma} 
        \int_0^\infty dq_\gamma q^2_\gamma  
  Z_{I;\tilde K_\alpha,I';\tilde K_ \gamma}^{J^\pi T}
(q_\alpha,q_\gamma;E)
\cr                      
&& \times  \tau_{l_\gamma {j}_\gamma} (E-\epsilon_\gamma(q_\gamma))
                \  f_{\tilde K_\gamma}^{J^\pi T}  (q_\gamma) ,
\label{AGS}
\end{eqnarray}
where the integral kernel $ Z_{I;K_\alpha,I';K_ \beta}^{J^\pi T} $ 
is defined by
\begin{eqnarray}
&&Z_{I;\tilde K_\alpha,I';\tilde K_ \beta}^{J^\pi T} (q_\alpha,q_\beta;E) \equiv
\cr
&&  \bar \delta_{\alpha \beta} ~ \delta_{II'}  ~
\gamma_{In;l_\alpha,j_\alpha} ~ \gamma_{I'n;l_\beta,j_\beta}
\cr &&
\times
           \langle g_{I;l_\alpha {j}_\alpha}; q_\beta K_\beta J T |
  G_0 ^{(I)} | g_{I';l_\beta {j}_\beta} ;q_\beta K_\beta J T \rangle ~~
 \label{eq:Z}
\end{eqnarray}
and   $\epsilon_\gamma (q_\gamma) $  is  
 $ {q_\gamma^2 \over 2 \mu_\gamma }$, and $E$ is a 
total energy of the three-body c.m. system.
Eq. (\ref{eq:Z}) is only changed with the parts of $\delta_{II'}$ and 
$\gamma$ from Eq. (13) of \cite{Eskan1992}.
In addition, the free three-body Green's function $G_0$ can be written as
\begin{eqnarray}
&&G_0^{(G)} \equiv
 \langle G| \hat G_0 | G \rangle = {1 \over E- p_\alpha ^2 / (2 \nu_\alpha) 
-q_\alpha^2 /(2 \mu_\alpha) +i\epsilon}, \cr
&& G_0^{(X)} \equiv 
\langle X | \hat G_0 | X \rangle 
\cr &&= 
{ 1 \over E+\delta m- p_\alpha ^2 / (2 \nu_\alpha) 
-q_\alpha^2 /(2 \mu_\alpha) +i \epsilon},
\end{eqnarray}
where the reduced mass $\nu_\alpha$ and $\mu_\alpha$ are $m_\beta m_\gamma
/(m_\beta+m_\gamma)$ and 
$m_\alpha (m_\beta +m_\gamma)/(m_\alpha+m_\beta+m_\gamma) $, respectively.

In order to find out the three-body bound state or
resonance state we regard the AGS equations of Eq. (\ref{AGS}) 
as the eigen value equation 
\begin{eqnarray}
\eta \vec \psi = {\cal K}(E) \vec \psi 
\end{eqnarray}
where $\eta$ and ${\cal K}(E)$ are the eigen value and the integral 
kernel $Z(E) \tau $ in Eq. (\ref{AGS}), respectively.
We need to search for $E$ under a constraint $\eta =1$. 
Our basic technique is based on the Gauss -
Seidel method to solve the eigen value equation. 
The typical iteration of the procedure is a few hundred times to
reach the stable solutions.
Performance of the integral for the complex momentum $q_\gamma$ 
takes the integral pass  as well as 
2-body momentum $p$ shown in Fig. \ref{Fig2}.
The contour deformation angle $\theta $ is defined
\begin{eqnarray}
p_{\rm complex}\equiv p \exp (-i \theta), q_{\rm complex} \equiv q \exp (-i \theta) 
 \end{eqnarray}
The accuracy of the calculation is sufficiently saved within $ \theta \le
{ \pi \over 3 }$.

\section{III. Numerical results}
We applied the above-mentioned scheme to the core-nucleus+2n systems of
$^8$He and $^{10}$He.
The results of these systems are separately  
demonstrated in the next subsections.

\subsection{$^8$He nucleus}

We treat, here, $^8$He as the $^6$He-n-n three-body system.
The energy shift $\delta m$ between the ground state $G$ and
the first excited state $X$ of the core-nucleus $^6$He is
1.8 MeV.  
There are low-lying three resonance states in $^7$He, which are submitted 
 J$^\pi$=$({3\over2})^-$ (g.s.;$\Gamma_{cm}$=0.150$\pm$0.020 
MeV\cite{Tilley:2002vg}),
J$^\pi$=$({1 \over 2})^-$ ($E_x=$0.9$\pm$0.5 MeV,  
$\Gamma_{cm}$=1.0$\pm$0.9 MeV\cite{Skaza:2006cc} ) and 
J$^\pi$=$({5\over2})^-$  ($E_x=$2.92 $\pm$ 0.09 MeV, 
$\Gamma_{cm} $=1.990 $\pm$ 0.170 MeV)\cite{Tilley:2002vg}. 
The energy level of the 
ground state is 0.445 MeV \cite{Tilley:2002vg} 
from the threshold of $^6$He and neutron,
we have each $E_2^{res.} $ in Table \ref{table1}. 
Using these experimental data we list the coupling constants $\gamma_I^2$
obtained by solving our model equations (\ref{cond}). 
For the sake of simplicity the reduced mass
$\nu$ is ${6\over 7}m_N$ with nucleon mass $m_N=$939 MeV.

 \begin{table}
 \caption{\label{table1}
Parameters for  $^6$He(0$^+$)-n + $^6$He(2$^+$)-n potential. 
 The resonance energies are 
measured from the $^6$He+n threshold. The strengths $\gamma^2$
are in unit of fm$^{-5}$ for P wave, and fm$^{-7}$ for F wave. The parameters 
$\beta_G$ and $\beta_X$ are commonly taken 1.5166 fm$^{-1}$. ($\beta _{nn} $= 1.1648 fm$^{-1}$) }
 \begin{ruledtabular}
 \begin{tabular}{cccccc}
    $E_2^{res}$ [MeV]  & partial wave  &$l_G$&$\gamma_G^2$ & $l_X$& $\gamma_X^2$   \\ \hline
 0.445 -- i 0.075  \cite{Tilley:2002vg}  & $^2$P$_{3/2}$+$^{4,6}$P$_{3/2}$ & 1 & 4.1655 &1 & 6.1580  \\
 1.345 -- i 0.5  \cite{Meister:2002zz,Skaza:2006cc}  & $^2$P$_{1/2}$+$^{4}$P$_{1/2}$ & 1 & 5.3966 &1 & 4.0418  \\
 3.37 --i 0.995  \cite{Tilley:2002vg} & $^2$F$_{5/2}$+$^{4,2}$P$_{5/2}$ & 3 & 116.80 &1 & 7.6144 \\ \hline
   nn channel & $^1$S$_0$ & 0&  0.3943 & 0 & 0.3943
 \end{tabular}
 \end{ruledtabular}
 \end{table}

The possible quantum numbers of  3-body partial wave  of J$^\pi$=0$^+$
 are listed  in table \ref{tbHe8_1}. 
There are 10 channels for J$^\pi$=0$^+$ ground 
state of $^8$He, and
32 channels for J$^\pi$=2$^+$.
 In table \ref{tbhe8} our theoretical predictions are demonstrated with the 
recent experimental data.
 Energy levels are reasonably well obtained to describe the data, however,
there is a tendency of large width.


 \begin{table}
 \caption{\label{tbHe8_1} Set of the quantum numbers for 
J$^\pi$=0$^+$ state of $^{8}$He nucleus. 
The quantum numbers for the particle channel
$\alpha$=3 is obtained from $\alpha$=1 by only cyclically label 
replacing $ s_\alpha \to
s_\beta \to s_\gamma \to s_\alpha$. }
 \begin{ruledtabular}
 \begin{tabular}{cccccccccccc}
$K_\alpha$ &$\tilde K_\alpha$
&$\alpha$&I& ${\cal L}_\alpha$ & ${\cal S}_\alpha$ &
$j_\alpha$ & $l_\alpha $& $S_\alpha$ &$s_\alpha$& $s_\beta$ & $s_\gamma$   \\ \hline
  1 &1&1&G&1 &   1  &  3/2& 1 &   1/2&    1/2 &   1/2 &   0    \\
  2 &1&1&X&1 &   1  &  3/2& 1 &   3/2&    1/2 &   1/2 &   2    \\
  3 &1&1&X&0 &   0  &  3/2& 1 &   5/2&    1/2 &   1/2 &   2    \\
  4 &2&1&G&0 &   0  &  1/2& 1 &   1/2&    1/2 &   1/2 &   0    \\
  5 &2&1&X&0 &   0  &  1/2& 1 &   3/2&    1/2 &   1/2 &   2    \\
  6 &3&1&G&0 &   0  &  5/2& 3 &   1/2&    1/2 &   1/2 &   0    \\
  7 &3&1&X&0 &   0  &  5/2& 1 &   3/2&    1/2 &   1/2 &   2    \\
  8 &3&1&X&0 &   0  &  5/2& 1 &   5/2&    1/2 &   1/2 &   2    \\
  9 &4&2&G&0 &   0  &  0 &   0&    0 &     0  &   1/2 &   1/2   \\
 10 &5&2&X&2 &   2  &  0 &   0&    0 &     2  &   1/2 &   1/2   
 \end{tabular}
 \end{ruledtabular}
 \end{table}  

 \begin{table}
 \caption{\label{tbhe8}
The predicted energy levels of $^8$He nucleus from $^6$He + n + n threshold. 
The resonance Energy $E$ equals to $E^{(r)} - i \Gamma /2$.   Unit is in MeV.}
 \begin{ruledtabular}
 \begin{tabular}{ccccccc}
   \multicolumn{1}{c}{J$^\pi$} &  \multicolumn{2}{c}{present work}  &  
 \multicolumn{2}{c}{Exp.}  \\ 
 \cline{2-3} \cline{4-5} \\
             & $ E^{(r)} $  &  $\Gamma$ &   $E^{(r)}$   & $ \Gamma$   \\ \hline
  0$^+$     &       -1.35  &           &    -2.14      &             \\
  2$^+$     &         2.01 &    2.12  &     1.06 $\pm$ 0.5      &  0.6 $\pm$ 0.2
 \end{tabular}
 \end{ruledtabular}
 \end{table}

\subsection{$^{10}$He nucleus}

The  $^{10}$He nucleus is here treated as the $^8$He-n-n three-body system.
The energy shift $\delta m$ between the ground state $G$ and
the first excited state $X$ of the core-nucleus $^8$He is
3.1 MeV.
There are low-lying two resonance states in $^9$He, which are submitted
 J$^\pi$=$({1\over2})^-$ (g.s.;$\Gamma_{cm}$=0.10$\pm$0.06 MeV)\cite{Tilley:2004zz}
and
J$^\pi$=$({1 \over 2})^+$ ($E_x=$1.15$\pm$0.10 MeV,
$\Gamma_{cm}$=0.7$\pm$0.2 MeV)\cite{Tilley:2004zz}.
The energy level of the
ground state is 1.27  MeV \cite{Tilley:2004zz}
from the threshold of $^8$He and neutron,
we have each $E_2^{res.} $ in Table \ref{table2}.
Using these experimental data we obtained the coupling constants $\gamma_I^2$
by our model equations (\ref{cond}) as well as the case of $^8$He.
Because of simplicity the reduced mass
$\nu$ is ${8\over 9}m_N$.

 \begin{table}
 \caption{\label{table2}
Parameters for  $^8$He(0$^+$)-n + $^8$He(2$^+$)-n potential.
 The resonance energies are
measured from the $^8$He+n threshold. The strengths $\gamma^2$
are in unit of fm$^{-5}$ for P wave, and fm$^{-7}$ for F wave. The parameters
$\beta_G$ and $\beta_X$ are commonly taken 1.5166 fm$^{-1}$.}
 \begin{ruledtabular}
 \begin{tabular}{cccccc}
 $E_2^{res}$ [MeV]  & partial wave  &$l_G$&$\gamma_G^2$ & $l_X$& $\gamma_X^2$   \\ \hline
 1.27 -- i 0.05 \cite{Tilley:2004zz}  & $^2$P$_{1/2}$+$^{4}$P$_{1/2}$ & 1 & 
0.44601 &1 &10.181  \\
 2.42 -- i 0.35 \cite{Tilley:2004zz}  & $^2$S$_{1/2}$+$^{4,6}$D$_{1/2}$ & 0 & 
0.016538 &2 & 118.42  \\
 \end{tabular}
 \end{ruledtabular}
 \end{table}

The possible quantum numbers of  3-body partial wave  of J$^\pi$=0$^+$
 are listed  in table \ref{tbHe10}.
There are 7 channels for J$^\pi$=0$^+$ ground
state of $^{10}$He, and
7 channels for J$^\pi$=2$^+$.
 In table \ref{tbhe10} our theoretical predictions are demonstrated with the
recent experimental data.
The state (1$^-$) not found in the experiment is obtained.
Although we would like to recommend to measure it, the clustering of the 
state may be not well developed.

\begin{table}
 \caption{\label{tbHe10} Set of the quantum numbers for J$^\pi$=0$^+$ state of $^{10}$He nucleus}
 \begin{ruledtabular}
 \begin{tabular}{cccccccccccc}
$K_\alpha$ &$\tilde K_\alpha$&
$\alpha$&I& ${\cal L}_\alpha$ & ${\cal S}_\alpha$ &
$j_\alpha$ & $l_\alpha $& $S_\alpha$ &$s_\alpha$& $s_\beta$ & $s_\gamma$   \\ \hline
  1 &1&1&G&1 &   1  &  1/2& 1 &   1/2&    1/2 &   1/2 &   0    \\
  2 &1&1&X&1 &   1  &  1/2& 1 &   3/2&    1/2 &   1/2 &   2    \\
  3 &2&1&G&0 &   0  &  1/2& 0 &   1/2&    1/2 &   1/2 &   0    \\
  4 &2&1&X&0 &   0  &  1/2& 2 &   3/2&    1/2 &   1/2 &   2    \\
  5 &2&1&X&0 &   0  &  1/2& 2 &   5/2&    1/2 &   1/2 &   2    \\
  6 &3&2&G&0 &   0  &  0 &   0&    0 &     0  &   1/2 &   1/2   \\
  7 &4&2&X&2 &   2  &  0 &   0&    0 &     2  &   1/2 &   1/2
 \end{tabular}
 \end{ruledtabular}
 \end{table}

 \begin{table}
 \caption{\label{tbhe10}
The predicted energy levels of $^{10}$He nucleus form  $^8$He + n + n threshold.
The resonance Energy $E$ equals to $E^{(r)} - i \Gamma /2$.   Unit is in MeV.}
 \begin{ruledtabular}
 \begin{tabular}{cccccc}
   \multicolumn{1}{c}{J$^\pi$} &  \multicolumn{2}{c}{present work}  &
 \multicolumn{2}{c}{Exp.}  \\
 \cline{2-3} \cline{4-5} \\
             & $ E^{(r)} $  &  $\Gamma$ &   $E^{(r)}$   & $ \Gamma$   \\ \hline
  0$^+$     &    0.803 &  0.665  & 1.069   &  0.3 $\pm$ 0.2  &             \\
  1$^-$     &     1.25 &  0.21   &      &         &             \\
  2$^+$     &    3.97  &  4.71   & 4.31 $\pm$ 0.20    & 0.6$\pm$0.3   &  
 \end{tabular}
 \end{ruledtabular}
 \end{table}


\section{IV. CONCLUSION}

We have been conducting research on $^{6,8,10}$He isotopes based on the 
three-cluster model. 
Incorporating the core-nucleus excitation we deal with double
Hilbert spaces. 
In the sense of ab initio calculation 
 only from the fundamental NN potential  double 
Hilbert spaces are not necessary.
The three-cluster model requires effective cluster 
potential between the core-nucleus
and neutron. 
Even though the potential made by the sufficient data in each space, it is
not always necessarily useful in the three-cluster model.
We have adopted a separable potential of rank 1, which bounds both of Hilbert
spaces. Coupling constants in the two spaces can be determined by its width and
the energy level of the resonance state in subsystem. 

\begin{figure}[htbp]
\begin{center}
\includegraphics[scale=0.4 ,angle=-0]{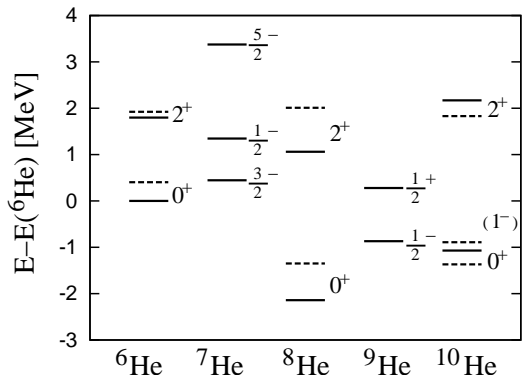}
\end{center}
\caption{
Energy levels of He isotopes normalized to the $^6$He ground state energy.
The dashed lines are corresponding to our theoretical predictions.
The solid lines are taken from experimental data 
\cite{Tilley:2002vg,Tilley:2004zz,Skaza:2006cc}.
 }
\label{LEVEL}
\end{figure}

There are the ground 0$^+$ and the excited 2$^+$ states in both of
$^8$He and $^{10}$He.
In Fig. \ref{LEVEL} their energy levels are shown.
The solid (dashed) level lines are corresponding to experimental data 
(theoretical predictions).
The energy level of $^6$He are obtained from \cite{Eskan1992} which are recalculated 
to check our program code. Our numbers of $^6$He agree with \cite{Eskan1992}.
The states of $^8$He and $^{10}$He fairly appear as our theoretical prediction.
Comparing with the case of $^{10}$He, we obtain rather a large difference ($\approx 
$1 MeV)   
between data and prediction in $^8$He. 
The level 1$^-$ is found, which is close to the 0$^+$ state. However,
this might be a simple spurious state because
the real state of 1$^-$ may not be a cluster state.
Expected theoretical decay width does not reproduce the experiment so much as a whole.

Although it is difficult to evaluate the accuracy of our model only 
by having investigated about a few nuclei, we would like to mention 
that our results were reasonably satisfactory.
For the sake of proving the effectiveness of our model 
we can only continue to predict unknown states 
which are not measured yet.

\begin{acknowledgments}

We would like to thank Prof. Yasuro Koike (Hosei University) and
Prof. Susumu Shimoura (CNS, RIKEN) for helping us
via fruitful discussion. One of authors (M.Y.) acknowledges
the support of the Theory Group of the Research Center for
Nuclear Physics (RCNP) at Osaka University. The numerical
calculations were performed on the interactive server at RCNP,
Osaka University, on the supercomputer cluster of the JSC,
J\"ulich, Germany, and in part on High Performance Computing
System at Tokyo University of Science.
\end{acknowledgments}

\bibliography{Heiso2}
\end{document}